# HoloStain: Holographic virtual staining of individual biological cells

Yoav N. Nygate[1], Mattan Levi[1], Simcha K. Mirsky[1], Nir A. Turko[1], Moran Rubin[1], Itay Barnea[1], Gili Dardikman-Yoffe[1], Alon Shalev[2], Natan T. Shaked[1,*]

[1] *Tel Aviv University, Faculty of Engineering, Department of Biomedical Engineering, Tel Aviv, Israel*
[2] *QART Medical, Ra'anana, Israel*
*\* Corresponding author: nshaked@tau.ac.il*

**Many medical and biological protocols for analyzing individual biological cells involve morphological evaluation based on cell staining, designed to enhance imaging contrast and enable clinicians and biologists to differentiate between various cell organelles. However, cell staining is not always allowed in certain medical procedures. In other cases, staining may be time consuming or expensive to implement. Furthermore, staining protocols may be operator-sensitive, and hence lead to varying analytical results by different users, as well as cause artificial imaging artifacts or false heterogeneity. Here, we present a new deep-learning approach, called HoloStain, which converts images of isolated biological cells acquired without staining by holographic microscopy to their virtually stained images. We demonstrate this approach for human sperm cells, as there is a well-established protocol and global standardization for characterizing the morphology of stained human sperm cells for fertility evaluation, but, on the other hand, staining might be cytotoxic and thus is not allowed during human in vitro fertilization (IVF). We use deep convolutional Generative Adversarial Networks (DCGANs) with training that is based on both the quantitative phase images and two gradient phase images, all extracted from the digital holograms of the stain-free cells, with the ground truth of bright-field images of the same cells that subsequently underwent chemical staining. After the training stage, the deep neural network can take images of unseen sperm cells, retrieved from the coinciding holograms acquired without staining, and convert them to their stain-like images. To validate the quality of our virtual staining approach, an experienced embryologist analyzed the unstained cells, the virtually stained cells, and the chemically stained sperm cells several times in a blinded and randomized manner. We obtained a 5-fold recall (sensitivity) improvement in the analysis results, demonstrating the advantage of using virtual staining for sperm cell analysis. With the introduction of simple holographic imaging methods in clinical settings, the proposed method has a great potential to become a common practice in human IVF procedures, as well as to significantly simplify and facilitate other cell analyses and techniques such as imaging flow cytometry.**

Digital pathology and cytology are emerging fields that are eventually expected to become fully automated and non-subjective, with applications ranging from routine clinical tests of body fluids to more complex biological research. Part of these analyses is based on morphological evaluation of individual cells. Cells in vitro are mostly transparent under regular light microscopy, and therefore cannot be imaged well without external stains or contrast agents. However, cell staining is time consuming and the staining materials might be harmful to the cells, resulting in the prohibition of chemical staining in certain medical procedures. Specifically, cell staining is not allowed during the selection of sperm cells for human in vitro fertilization (IVF), not allowing the ability of intracellular morphology evaluation. Since off-axis holography records the quantitative phase profile, which takes into account the cell refractive index and physical thickness, it can obtain the quantitative topographic maps of the cell from a single camera exposure. This method creates great imaging contrast without the need for external contrast agents. Even when using contrast agents in imaging flow cytometry, the fact that phase profile is quantitative and accounts for the cell internal refractive indices gives rise to new parameters with medical relevance that have not been available in flow cytometry before, such as the dry mass of the cells[1,2]. Until recently, holographic cell imaging could not be implemented in clinical settings due to the bulkiness and non-portability of the optical system, as well as the requirement for specific optical skills to align and use it. In the last years, successful efforts have been made to make these wavefront sensors affordable to clinical use[3,4]. Our approach, called interferometric phase microscopy (IPM), is based on the usage of microscopes already existing in medical clinics and attaching a portable interferometric module to their exit port[4]. This wavefront sensor is compact, inexpensive and easy to operate, making this technology accessible and affordable to clinicians' direct use. However, in spite of this technique potential to aid in cell analysis, existing and well-established protocols for morphological cells evaluation are still based on chemical staining of the cell organelles, rather than on the topographic maps obtained by holography. Thus, in spite of its potential, digital holography is far from full integration into medical procedures and biological protocols.

In this paper, we propose a new deep learning approach for transforming quantitative phase maps of individual biological cells extracted from digital holograms to their virtually stained images, which are very similar to their actual chemical staining images. We have chosen to demonstrate this technique for stain-free sperm imaging, since there is an established World Health Organization (WHO) protocol for morphological evaluation of sperm cells during fertility evaluation. However, this protocol cannot be fully implemented in human IVF procedures due to the prohibition of using cell staining.

In the past several years, deep learning has emerged as a beneficial tool in the medical imaging field, simplifying many complex image analysis tasks[5]. Deep learning enables the computer to learn specific tasks based on observed data. This is done by feeding the data through many processing layers, which, after a training procedure, are able to estimate complex data representations[6]. In medical imaging, deep learning was already demonstrated as a beneficial method for performing segmentation of medical images[7–9] and solving different inverse problems in the medical imaging field[10]. Generative Adversarial Networks (GANs) is a deep learning framework, which allows the training of generative models by performing an adversarial process between two deep learning networks, a generator model and a discriminator model[11]. In particular, deep convolutional GANs (DCGANs) have been shown successful for training generative models for image generation tasks[12,13].

Lately, the combination of holographic imaging and deep learning for classifying between different types of biological cells has been shown successful[14,15]. Furthermore, recently several deep learning frameworks were used for performing virtual histology of biological tissue sections from auto-fluorescence signals and from quantitative phase images that were reconstructed from lens free in-line holograms[16,17]. However, to the best of our knowledge, virtual staining of individual biological cells, rather than full tissue sections, was never been shown successful based solely on stain-free holographic imaging, not allowing its use for many medical and biological procedures, such as imaging flow cytometry and sperm selection for IVF. Our method, named HoloStain, uses DCGANs to transform quantitative phase images and phase gradient images, extracted from stain-free digital holograms, to their stain-based versions that are similar to the conventional chemically stained images, making holographic imaging much more relevant for direct clinical use.

# Results

## Virtual staining of sperm cells

We acquired 166 human sperm cells without staining using off-axis digital holographic microscopy, and then acquired the same cells after staining them by QuickStain using a conventional bright-field microscope. All images were acquired with $60\times$ oil-immersion microscope objective. The optical system details are presented in the Methods Section. Next, we used image augmentation to create an 8-fold increase in the dataset size. Overall, our dataset contained 1328 image pairs of stain-free off-axis holograms of sperm cells, and their stain-based bright-field image counterparts. Each of the stain-free holograms was used to extract three images: a quantitative phase image and two synthetic gradient phase images in two orthogonal directions (see Method Section for the digital processing used). These additional phase gradient images were necessary for the success of the virtual staining process (thus, the quantitative phase images were not enough for the network convergence). Overall, for each cell, we had a batch of four images: stain-free quantitative phase image, two stain-free phase gradient images, and chemical staining as the ground truth. We divided the data to 1100 batches for training and 228 batches for testing. We then constructed a DCGAN model for obtaining virtual staining. The DCGAN framework, which is constructed from a generator network and a discriminator network competing with each other, was first trained on the 1100 batches of sperm cells. The generator network receives as an input a batch of quantitative phase images and the two gradient phase images and outputs the generated virtually stained image. The discriminator network is trained to distinguish between the generated and the chemically stained images. It first receives both the generator input batch with the chemically stained image, and then receives the generator input batch with the generated output. By balancing between the loss functions of the generator and the discriminator, the generator is trained to create the correct virtually stained image. The full networks' architectures are given in the Methods Section.

After training, the DCGAN model was tested on the 228 batches that were never seen by the model before. In this case, the generator was used in order to create the virtually stained images of the sperm cells, where the coinciding stain-based bright-field images were used for calculating a similarity metric between the real and generated images.

Figure 1 presents examples of the results obtained by HoloStain on several sperm cells from the test dataset, never seen by the networks in the training step, having normal and pathological morphologies. Figure 1(a) shows the stain-free off-axis holograms of the cells. Figures 1(b-d) show the coinciding quantitative phase images and gradient phase images, directly extracted from the holograms shown in (a). This triplet of stain-free images is the input to the previously trained generator network. Figure 1(e) shows the generated virtually stained images, the outputs of the trained generator network. Figure 1(f) shows the chemically stained bright-field image of the coinciding cells, for comparison. The resulting virtually stained images in Figure 1(e) have a similar color scheme to that of the chemically stained images in Figure 1(f). In addition, it can be seen that noise and debris are eliminated by the HoloStain method, resulting in a clean and even background surrounding the sample, which further eases the morphological examination of the cell.

For each of the 228 test images, the mean average error (MAE, see methods section) was calculated between the virtually stained image and the chemically stained image, resulting in an overall average MAE of $0.1566 \pm 0.0446$.

In holographic imaging, the whole complex wavefront can be reconstructed from the captured holograms, allowing it to be propagated such that unfocused objects will come into focus. Thus, using HoloStain, we can now present the virtually stained images even if the cells were out of focus during acquisition, which can help in increasing the acquisition throughput in comparison to bright-field imaging, even if the cells are chemically stained. Often, when imaging a certain population of sperm cells, the clinician would need to constantly change the focus of the microscope in order to view all of the cells present. Using HoloStain, a single hologram can be captured with out-of-focus cells. Then, by reconstructing the whole complex wavefront, each sperm cell can be propagated into focus and then virtually stained. Figure 2 demonstrates an out-of-focus cell that is brought into focus by propagating the reconstructed complex wavefront (see Methods Section for details) and then virtually stained by HoloStain method.

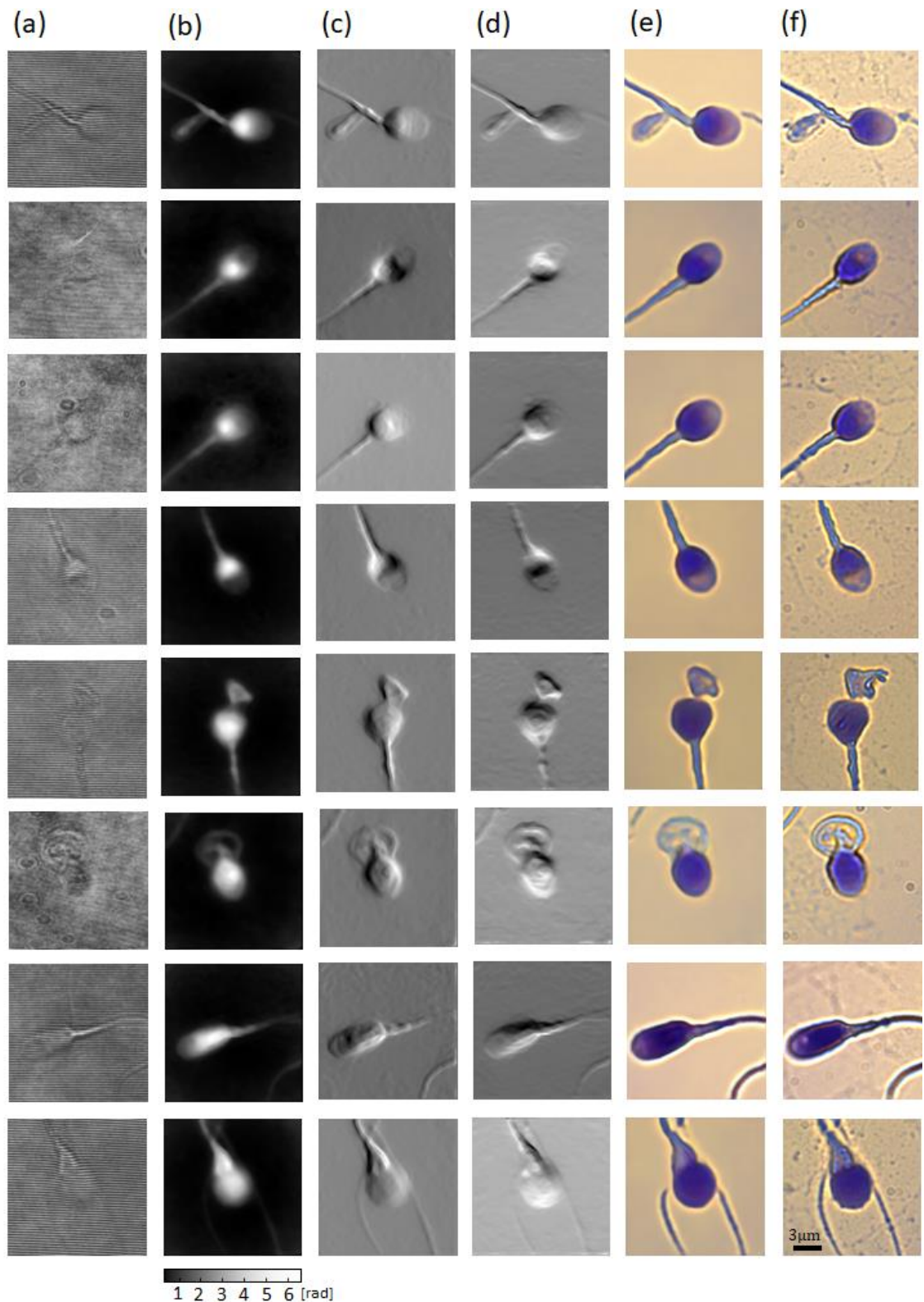

(a)  (b)  (c)  (d)  (e)  (f)

1 2 3 4 5 6 [rad]

3μm

**Fig. 1 | Examples of HoloStain results for individual sperm cell imaging.** The first four rows show normal morphology cells. The last four rows show pathological cells. **a**, off-axis image holograms of the cells acquired without staining. **b**, the coinciding quantitative phase images extracted from the holograms. **c**, the coinciding horizontal phase gradients extracted from the holograms. **d**, the coinciding vertical phase gradients extracted from the holograms. **e**, the coinciding virtual stained images, generated by the generator network, where (b-d) are the input to the generator. **f**, the coinciding bright-field chemically stain images of the same cells, for comparison.

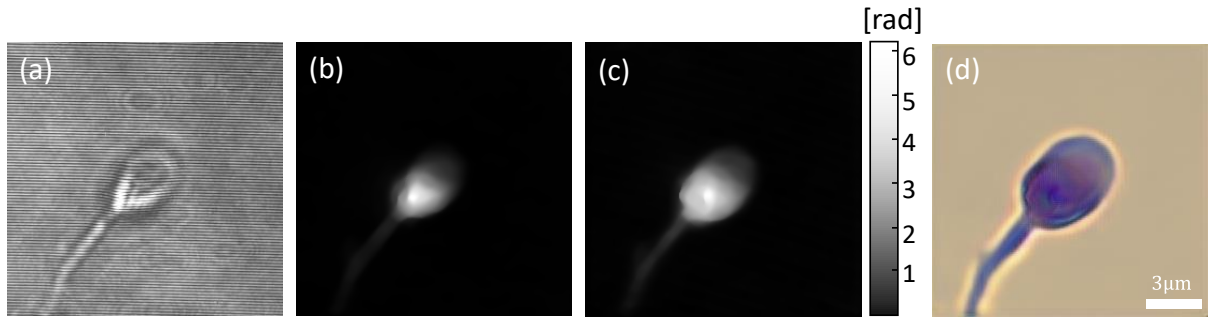

**Fig. 2 | Virtual staining of a sperm cell propagated into focus. a**, hologram of an out-of-focus sperm cell. **b**, phase reconstruction of the out-of-focus sperm cell. **c**, phase reconstruction after propagating the complexed wavefront by −1.6 μm. **d**, virtual staining of the refocused sperm cell.

## Classification of the sperm cells

In order to assess the effectiveness of virtually stained sperm cells for performing sperm quality classification, five datasets of different sperm cells were created. The first dataset contains only bright-field image of sperm cells without staining, which is frequently the current practice in analyzing sperm cells today. The second database contains the bright field images of the respective sperm cells, but now with staining using QuickStain for sperm cells. The third dataset contains images of the respective sperm cells that have been acquired without staining using off-axis holography and have been virtually stained using HoloStain. The fourth dataset contains the stain-free quantitative phase images of the respective sperm cells. The fifth database contains one of the stain-free phase-gradient images of the respective cells, which resembles differential interference contract (DIC) images. Since there is no well-established automatic standard for sperm cell morphological evaluation, we asked an experienced embryologist to analyze each sperm image in each of the five datasets and classify it, separately and independently, as normal or abnormal, using the World Health Organization (WHO) criteria for sperm cell analysis. It is important to note that the datasets were presented to the embryologist in a randomized and blinded manner up to four times in order to minimize the effect of subjective analysis. Three confusion matrices were calculated. Abnormal sperm cells were classified as negative labels (0), normal sperm cells were classified as positive labels (1), and the chemically stained sperm cells were regarded as the ground-truth labels.

For performing IVF procedures, where the selection of healthy sperm cells is considered critical, high precision for positive labels is required, where precision is defined by Eq. (1) below.

Moreover, when the selection of several healthy sperm cells is needed, high recall is required as well, where recall (also called sensitivity) is defined by Eq. (2) below. Overall, an F1 score, defined by Eq. (3) below, can be calculated in order to quantify the balance between the precision and recall of the classified cells in each dataset. These three metrics are mathematically defined as follows:

$$Precision = \frac{TP}{TP+FP}, \tag{1}$$

$$Recall = \frac{TP}{FN+TP}, \tag{2}$$

$$F1 = 2 \times \frac{Precision \times Recall}{Precision + Recall}, \tag{3}$$

where TP signifies true positives – cells which are classified as positive and their corresponding chemically stained cells are classified as positive as well, FP signifies false positives – cells which are classified as positive and their corresponding chemically stained cells are classified as negative, TN signifies true negatives – cells which are classified as negative and their corresponding chemically stained cells are classified as negatives as well, and FN signifies false negatives – cells which are classified as negative but their corresponding chemically stained cells are classified as positive.

From the confusion matrices in Fig. 3(a), a precision of 1.0 was calculated across all datasets. This indicates that the embryologist was very conservative in classifying the cells, since he did not classify unhealthy cells as healthy cells in all methods in comparison to the chemical staining, even based on the stain-free bright field images only. This indicates that his classification efficiency was low, since it would take him longer to choose good cells, especially in cases of pathologic sperm, where good cells are rare. Thus, virtual staining, providing contrast similar to chemical staining, is expected to make the embryologist's classification work in choosing healthy sperm cells less tedious by making him less conservative in classifying cells as healthy.

Furthermore, the recall gradually increased when advancing from stain-free bright-field imaging to virtual staining. The following recall values were calculated for the stain-free bright field, phase gradient, quantitative phase, and virtual staining images, respectively: 0.143, 0.143, 0.286, and 0.714. This shows that the virtual staining dataset enabled the detection of more normal-morphology sperm cells compared to the other datasets. Finally, the following F1 scores were calculated for the stain-free bright field, phase gradient, quantitative phase, and virtual staining images, respectively: 0.25, 0.25, 0.444, and 0.833. The F1 score signifies the overall accuracy of classifying healthy sperm cells by taking into account both the precision and recall. From the presented F1 scores, a gradual increase in classification performance can be seen, where out of the four stain-free methods analyzed, HoloStain enables classification results closest to the gold standard, the chemical staining method. A visualization of these metrics can be seen in Figure 3(b).

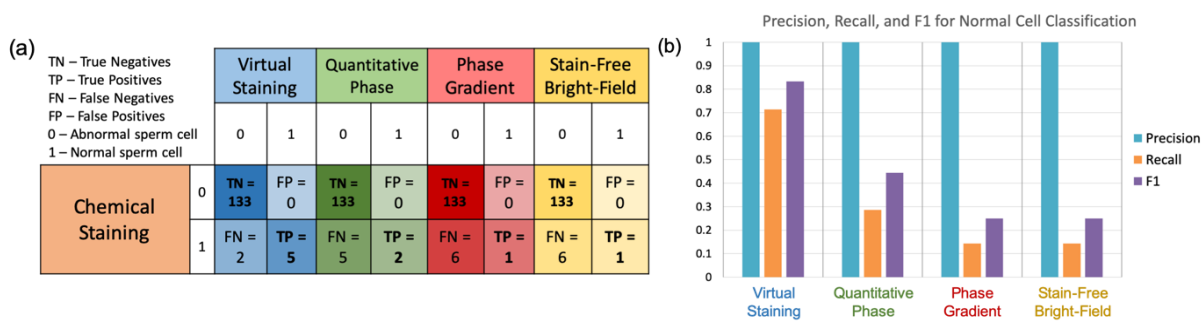

**Fig. 3 | Confusion matrices and metrics graph. a**, confusion matrices for the classification of virtually stained cells, phase image of the cells, synthetic phase gradient image of the cells, and the stain-free bright field image of the cells. **b**, comparison graph of the precision, recall and F1 metrics for the four datasets.

## Discussion and conclusion

The capability of virtually staining label-free biological samples has great potential for replacing conventional staining techniques of individual cells, including fluorescence and histochemical staining. Virtual staining saves preparation time, it is less prone to variability caused by different staining protocols and environmental conditions, and it provides a solution for circumstances where staining is prohibited. Yet, it gives the clinician or the biologist cell visualization similar to actual chemical staining, so that established protocols for diagnosis or research can be directly applied. Our deep learning-based technique, HoloStain, achieves virtual staining of quantitative phase images of individual biological cells acquired using a portable, clinic-ready off-axis IPM system that does not require cell staining. As demonstrated, HoloStain is able to generate images of sperm cells that are similar to the conventional staining method. Since reconstructing the full complex wavefront of imaged samples is possible using IPM systems, the virtual staining of out-of-focus cells can be generated as well. As a result, constantly focusing the microscope is not needed, simplifying the analysis process for the clinician and increasing the analysis throughput. We have demonstrated that analyzing the virtually stained sperm cells by an experienced embryologist achieves similar performance compared to the classification of the coinciding chemically stained sperm cells, where the latter is currently regarded as the gold standard for performing morphological analysis in sperm cells. Thus, we believe that HoloStain will provide a valuable tool for both researchers and clinicians alike for performing stain-free morphological analysis of biological cells, saving them valuable preparation time and allowing them to perform a more accurate analysis when chemically staining cells is prohibited or is too expensive to perform. Although we have demonstrated using HoloStain for sperm imaging, the same platform can be adapted in imaging other types of cells, paving the way for stain-free digital pathology and stain-free imaging flow cytometry.

## Methods

### Sample preparation and imaging of stain-free sperm cells

A drop of 5-10 µL of human sperm was smeared onto several clean microscopic slides with a 2x2 point grid painted onto them for localization of the sperm cells when transferring the samples between the systems. These smeared drops where then left to dry for 5 minutes and then fixed to the slides with 98% ethanol for 10 minutes. The slides where then imaged using an IPM system, which can be seen in Figure 4(a). This system consisted of the $\tau$-Interferometer connected at the output of an inverted microscope. A supercontinuum fiber light source (SC-400-4 Fianium) connected to an acousto-optical tunable filter (SC-AOTF, Fianium) was used as the light source for the inverted microscope, emitting wavelengths of $532 \pm 3.1$ nm. The beam first passes through the sample, then magnified using the microscope objective MO (63×, 1.4 NA, oil immersion, infinity-corrected) and passes through a spherical tube lens TL (150 mm focal length). Then, it passes through lens L1 (100 mm focal length), which Fourier transforms the beam, and beam splitter BS splits the beam into two separate beams. One beam passes straight through the beam splitter and then reflected back and shifted by retro-reflector RR. This beam is then reflected by the beam splitter and inverse Fourier transformed by lens L2 (150 mm focal length) onto a CMOS camera with $1280 \times 1024$ pixels (pixel size of 5.2 µm, DCC1545M, Thorlabs). This beam acts as the sample beam in this interferometric setup. The second beam is reflected by the beam splitter onto a mirror-pinhole configuration, PH and M3, which spatially filters the beam, thus erasing the sample information, creating the reference beam. This beam is then reflected back and passes through the beam splitter, where it is then inverse Fourier transformed by lens L2 and interferes with the sample beam on the camera. The final result is an off-axis interference pattern, which is then transferred to the computer for further digital analysis.

**Imaging of stained sperm cells**

After the sperm cells were imaged using the IPM system, they were stained using QuickStain (Biological Industries) and left to dry for 15 min. Then, using the 2x2 point grid, the field of views captured using the IPM system were located once again and imaged using a bright-field microscope (Axio Observer D1, Zeiss).

**Digital reconstruction of the holograms**

The off-axis interference pattern captured by the camera can be used to extract the complex wavefront. Shortly, this off-axis hologram is digitally Fourier transformed, resulting in a zero order and two high-order cross-correlation terms. Each cross-correlation term contains the complex wavefront of the sample, which allows the extraction of the cell quantitative phase information. This reconstruction process is illustrated in Figure 4(b). Therefore, one of the cross-correlation terms is digitally cropped and inverse Fourier transformed. Then, in case the sperm image is out of focus, a digital propagation algorithm is applied. Our propagation method of choice was the Rayleigh-Sommerfeld propagation of the angular spectrum[18]. Finally, the phase information was extracted from the argument of the resulting complex wavefront, which then underwent a 2D phase unwrapping algorithm[19]

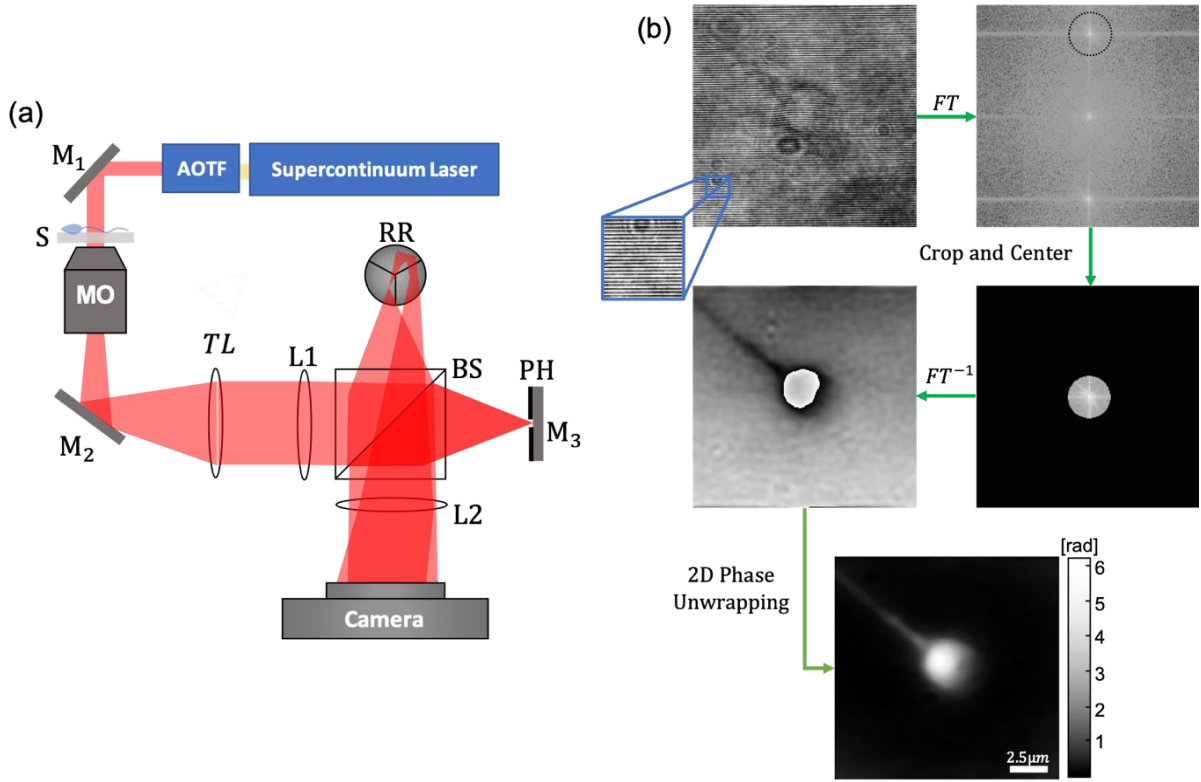

**Fig. 4 | Schematics of the optical setup and the reconstruction process. a**, the $\tau$-Interferometer positioned at the output of a commercialized microscope, which consists the following elements: a supercontinuum laser together with an acousto-optical tunable filter (AOTF) are used as the light source. M1, M2, and M3 are mirrors. S is the sample, MO is a microscope objective, and TL is the tube lens. L1 and L2 are lenses. PH is the pinhole, RR is the retroreflector, and BS is the beam splitter. **b**, the reconstruction process: the captured hologram is first Fourier transformed (FT), then one of the cross-correlation terms is cropped and centered, then it is inverse Fourier transformed ($FT^{-1}$) and the phase argument is extracted. Finally, the phase argument undergoes a 2D unwrapping algorithm.

## Calculation of synthetic phase gradient images from the phase images

In order to enhance high frequency information in the cell imaging, such as edges, and help with the training process, two synthetic phase gradient images were created from each phase image. Those images were generated by shifting the phase images by one pixel in one of the spatial directions ($x$ or $y$) and then subtracting the shifted image from the original phase image.

$$Grad\varphi_x = \varphi\,(x,y) - \varphi(x+1,y), \tag{5}$$

$$Grad\varphi_y = \varphi\,(x,y) - \varphi\,(x,y+1), \tag{6}$$

where $\varphi$ is the quantitative phase of the sample extracted from the off-axis hologram. The result of these phase gradients resembles to what can be obtained experimentally using a DIC microscope.

**Digital pre-processing**

As with the holograms, the bright-field images of the stained sperm cells were cropped into $256 \times 256$ pixels. This resulted in two datasets, one containing the bright-field images of the stained sperm cells, and the other containing the phase images and synthetic phase gradient images of the same sperm cells, where there is an exact overlap between the fields of view of the two datasets.

After constructing the above-mentioned dataset, it was further augmented by performing 90 degrees rotations for each image, and then horizontally flipping all existing and new images in the dataset. Overall, this caused the original dataset to increase by 8-fold.

**Training and testing procedures**

To train a deep learning model to virtually stain sperm cells, a DCGAN framework was used. This framework consisted of a generator network that was trained to create the virtually stained images from the stain-free quantitative phase and synthetic phase gradient images of the cells, and a discriminator network that was trained to discriminate between the generated and the chemically stained images.

As seen in Figure 5, in order to train the generator and discriminator networks, the generator receives an input batch $X$, which is a concatenation between the quantitative phase image and the two synthetic phase gradient images of the sperm cells, all extracted from the stain-free digital holograms of these cells. It is trained to generate $G$, which is the virtually stained image of the same sperm cell that was fed through the generator network. Since the discriminator is trained to distinguish between the generated and chemically stained images, in one case it receives $D_{XY}$, which indicates that the generator input $X$ is fed through the discriminator together with the chemically stained image of the sperm cell $Y$. In another case, the discriminator receives $D_{XG}$, which indicates that the generator input $X$ is fed through the discriminator together with the generated virtually stained image $G$.

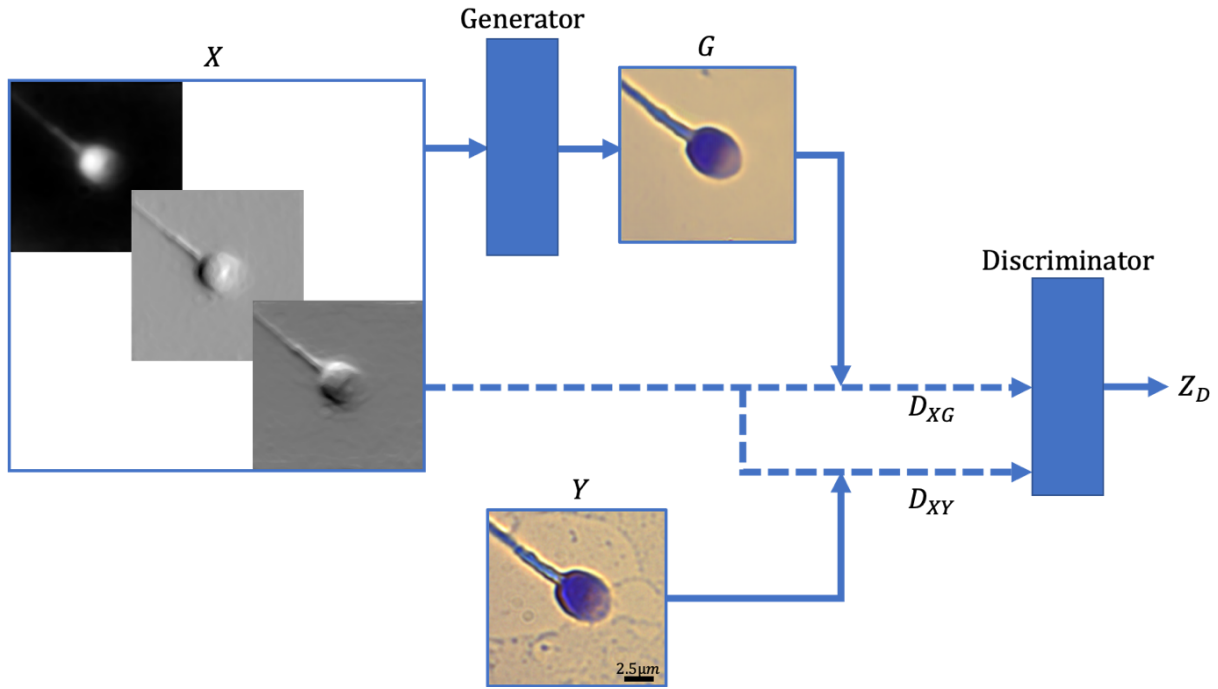

**Fig. 5 | Schematics of the training process.** *X* is the input to the generator network consisting of the stain-free quantitative phase image and two synthetic phase gradient images. *Y* is an image of the chemically stained sperm cell. The generator is trained to create *G*, the image of the virtually stained sperm cell. In one instance, *G* and *X* (marked as $D_{XG}$) are fed through the discriminator, which is trained to recognize this pair as fake images. In another instance, *Y* and *X* (marked as $D_{XY}$) are fed through the discriminator which is trained to recognize this pair as real images, while $Z_D$ is the output of the discriminator.

The losses for the networks use a combination of two error functions. The first is the mean absolute error (MAE), also known as $\mathcal{L}_1$ loss, which is calculated as follows:

$$\mathcal{L}_1(Y,G) = \frac{\sum_{i=1}^{n}|Y_i - G_i|}{n}. \tag{7}$$

The second is the sigmoid cross entropy (SCE), which is calculated as follows:

$$SCE(Z_D, Z) = max(Z_D, 0) - Z_D * Z + log\{1 + exp[-abs(Z_D)]\}, \tag{8}$$

where $Z_D$ is the output of the discriminator and $Z$ is the designated Boolean (1 for real images and 0 for fake images).

Overall, the generator loss is calculated using the following equation:

$$\mathcal{L}_G = \beta\mathcal{L}_1(Y,G) + SCE(Z_{XG}, 1), \tag{9}$$

where $Z_{XG}$ is the output of the discriminator when $D_{XG}$ is fed through it, $\beta$ is a multiplication factor used to give an emphasis on generating accurate virtually stained images, this value was set to $100$[13].

The discriminator loss is calculated as follows:

$$\mathcal{L}_D = SCE(Z_{XG}, 0) + SCE(Z_{XY}, 1). \tag{10}$$

where $Z_{XY}$ is the output of the discriminator when $D_{XY}$ is fed through it.

During the training stage, the generator loss and the discriminator loss were minimized using an ADAM optimizer[20]. In addition, the generator and discriminator use three types of activation functions. The first is a rectified linear unit (ReLU), which is calculated as follows:

$$ReLU(X) = max(X, 0). \tag{11}$$

The second is a leaky ReLU, which is calculated as follows:

$$LeakyReLU(X) = max(X, 0.2X). \tag{12}$$

The third is a sigmoid function, which is calculated as follows:

$$Sigmoid(X) = \frac{1}{1 + e^{-X}}. \tag{13}$$

Finally, also a hyperbolic tangent (tanh) is used, which can be calculated as follows:

$$tanh(X) = \frac{e^{2X} - 1}{e^{2X} + 1}. \tag{14}$$

**Internal architecture of the deep learning networks**

As seen in Figure 6, the generator network was based on a U-Net architecture[21]. This architecture consists of an encoder and a decoder with skip connections at every downsampling / upsampling stage. Each step of the encoder contains a convolutional block. Each convolutional block contains three sequences of a 2D convolution layer, a batch normalization layer[22], and a leaky ReLu activation function, as calculated in Eq. (12). The first and second convolutions in every step of the encoder are consisted of a convolutional layer with a kernel of 4 and a stride of 1, and the third block contains a convolutional layer with a kernel of 4 and a stride of 2. Overall, in each step of the encoder there is an increase of the depth of the filters by a factor of 2 and a decrease by a factor of 2 in the height and width dimensions. After the encoding step, nine residual net (ResNet) blocks were added in order to assist with the image transformation training of the generator[23-24]. The decoding stage consists of deconvolutions, a concatenation step for the skip connections, and two additional convolutional layers. The deconvolution step is made up of a sequence of a transpose 2D convolution layer with a kernel of 4 and a stride of 2, a batch normalization layer, and a ReLu activation function, as calculated in Eq. (11). This deconvolution step is followed by two sequences of a convolutional layer with a kernel of 4 and a stride of 1, a batch normalization layer, and a ReLu activation function. Overall, at each step of the decoder, the depth of the filters decreases by a factor of 2 and the height and width dimensions increase by a factor of 2. Furthermore, an additional skip connection is added at the final layer of the decoder, which performs an element-wise addition between the input image and the final layer of the generator, in order to decrease training time and achieve an image with a geometrical similarity to the input image[10].

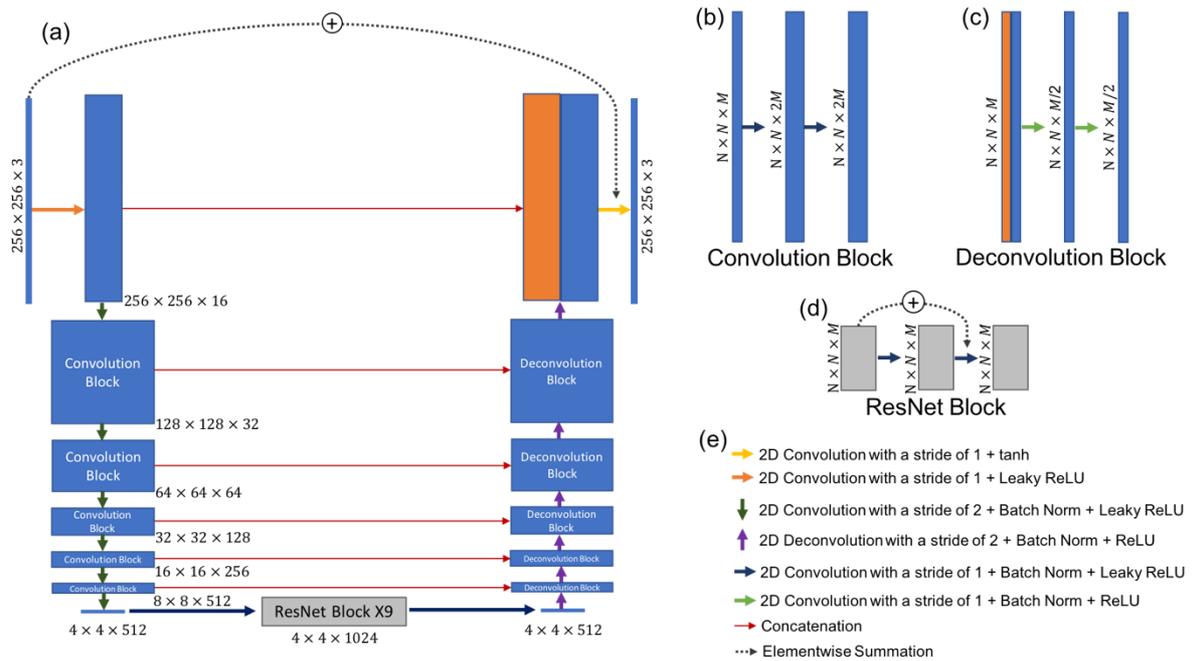

**Fig. 6 | Internal architecture of the generator network. a,** the overall architecture of the generator network. **b,** the inner architecture of a convolution block. **c,** the inner architecture of a deconvolutional block, where the orange rectangle signifies the concatenation process. **d,** the inner architecture of a residual net block. **e,** legend explaining the signification of each arrow in the architectures.

As seen in Figure 7, the discriminator model consists of convolutional blocks that are similar to the ones in the encoder step of the generator. As the input image passes along the discriminator, its depth is increased by a factor of 2 and the height and width dimensions are decreased by a factor of 2 until a $32 \times 32$-pixel image is created. The final two convolutional layers create a $30 \times 30$-pixel image with a depth of 1, and by applying the sigmoid function given in Eq. (13), each pixel in this image corresponds to the real or fake classification of overlapping patches within the input image. This framework was based on the PatchGAN discriminator[13], which increases training time and improves the sharpness of the generated images.

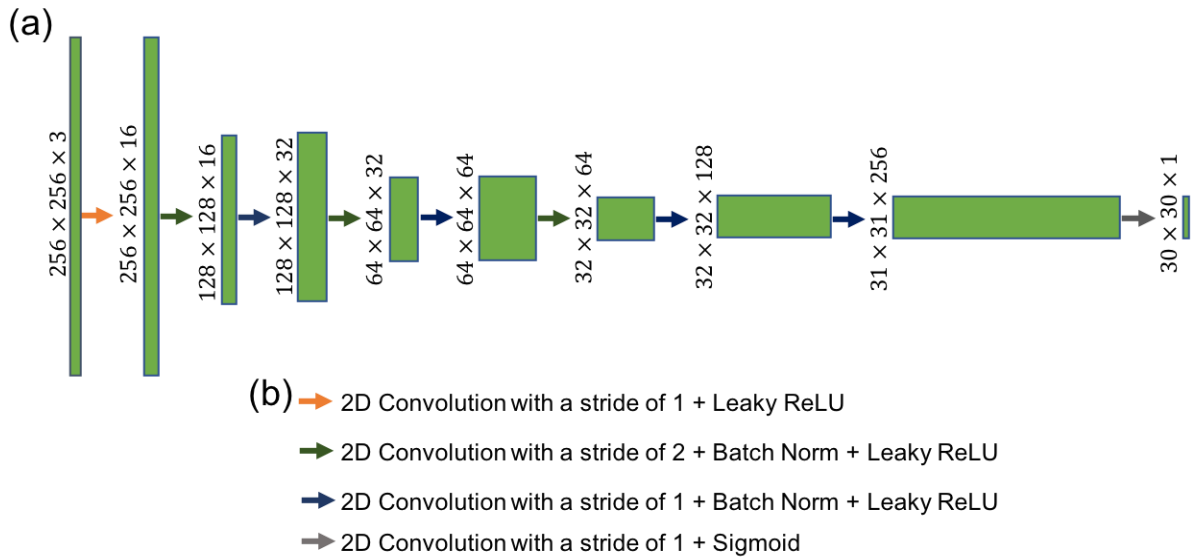

**Fig. 7 | Schematics of the discriminator network. a,** the overall architecture of the generator network. **b**, Legend explaining the signification of each arrow in the architecture.

## Implementation

The hologram reconstruction, synthetic phase gradient image calculation, and all of the digital pre-processing procedures that were performed on the images were implemented with MATLAB R2016b. All of the abovementioned processes were done on a desktop computer with an Intel Core i7-2600 CPU @ 3.40 GHz and 8.00GB RAM, running on a Windows 10 operating system (Microsoft). The deep learning architecture, and training/testing procedures were implemented in Python version 3.6.4 using the TensorFlow library version 1.10.0. The training and testing of the network was performed on a Tesla P100 GPU (NVIDIA) using the Google Cloud Platform. The framework was trained for 120 epochs, which lasted 31.5 hrs. Each image generation lasts approximately 0.08 seconds on a Nvidia Tesla P100 GPU.

**Data availability.** The data and code that support the results within this paper are available from the corresponding author upon reasonable request.

**Acknowledgments**

We thank the Horizon2020 ERC PoC grant of N.T.S. for funding this research.

**Author contributions**

N.T.S., A.S. and Y.N.N. conceived the idea. Y.N.N. scripted and formulated the HoloStain deep learning framework and analyzed all experimental data under the supervision and guidance of N.T.S. S.M. captured the microscopic data in this study. N.A.T. assisted with the digital processing of the acquired holograms. M.L. performed the morphological analyses of the chemically and virtually stained sperm cells. M.R. provided accessory code for this project and assisted with performing code review. I.B. performed all biological sample preparations. G.D. scripted the wavefront propagation algorithm. Y.N.N. and N.T.S wrote this manuscript. All authors discussed the results and commented on the manuscript.

**Competing interests**

The authors declare no competing interests.

**Additional information**

**Correspondence and request for materials** should be addressed to N.T.S.